\documentclass[journal]{IEEEtran}
\usepackage[dvips]{graphicx}

\usepackage[cmex10]{amsmath}

\usepackage[tight,footnotesize]{subfigure}

\begin{document}
%
\title{Longevity Studies in the CDF II Silicon Detector}
%
\author{Satyajit~Behari {\em on behalf of the CDF Collaboration} \\
        \IEEEmembership{The Johns Hopkins University,
        Baltimore, MD 21218.} \\
        E-mail: behari@fnal.gov}

\maketitle
\thispagestyle{empty}

%
%
\begin{abstract}
The CDF Run II silicon detector is the largest operating detector
of its kind in High Energy Physics, collecting $p\bar{p}$
collision data at the Fermilab Tevatron since 2001. It provides
precision tracking and vertexing which played a critical role in
the $B_s$ mixing discovery and is essential to the ongoing Higgs
Boson search and many other physics analyses carried out at CDF.
Due to the prolonged Tevatron Run II program the detector faces
unforeseen challenges while operating well beyond its design
parameters. Of particular concern is the radiation aging of the
silicon sensors which are expected to acquire $\sim$10 fb$^{-1}$
data, far above their design integrated luminosity of 2-3
fb$^{-1}$. In this paper we discuss the impact of radiation
damage to the sensors, their effect on the physics performance
and expectations for future operations of the two inner layers,
which have already inverted.
\end{abstract}


\IEEEpeerreviewmaketitle

%
%
\section{The CDF Run II Silicon Detectors}
\label{sec:det}
\IEEEPARstart{T}{he} Collider Detector at Fermilab (CDF)~\cite{CDF} 
is one of the
two multi-purpose particle physics detectors at the Tevatron
collider of Fermilab. The Tevatron has delivered an integrated
luminosity of 7 fb$^{-1}$ per experiment so far and is expected
to operate beyond 2010, delivering data in excess of 9 fb$^{-1}$.
The CDF Run II silicon detector provides precision tracking and
vertexing information essential for most of CDF's broad physics
program. It is also used in the CDF Silicon Vertex Trigger
(SVT)~\cite{SVT}, the first hardware-implemented displaced vertex
trigger at a hadron collider detector.

The silicon detector is comprised of three sub-detectors with a
total of 722k readout channels from 7 m$^2$ of silicon sensors.
A {\it ladder} is the smallest detector building block consisting
of 2-4 sensors, 4-16 readout chips and read out via a High
Density Interconnect (HDI).
The Layer 00 (L00)~\cite{L00} is installed directly on the beam
pipe. L00 uses radiation hard, single-sided $p$-in-$n$ sensors
from Hamamatsu at radius 1.62 cm and SGS Thomson at radius 1.35 cm
fabricated with design parameters similar to those in use by the
Compact Muon Solenoid (CMS)~\cite{CMS} silicon tracker at the LHC.
It also includes two oxygenated module made by Micron. The L00
sensors can be biased up to 500 V, limited by the power supply.
SVX-II~\cite{SVXII} is the core sub-detector and the only component
used in the SVT. SVX-II has 5 layers of double-sided $p$-in-$n$
sensors located at radii between 2.5 - 10.6 cm. The layers 0,1
and 3 are made of Hamamatsu sensors with the $n$ strips
perpendicular to the $p$ strips. The layers 2 and 4 are made of
Micron sensors with a 1.2$^o$ angle between the $n$ and $p$
strips. The Hamamatsu and Micron sensors can be biased up to
170 V and 70 V, respectively, limited by the breakdown voltage
of the integrated coupling capacitors and subtle sensor effects.
ISL~\cite{ISL} provides an extended forward tracking coverage
where CDF wire chamber has partial coverage. ISL has one central
layer at a radius of 22 cm and two forward layers at radii of
20 cm and 28 cm. They are made of double-sided $p$-in-$n$
Hamamatsu and Micron sensors with the $n$ strips at a stereo
angle of 1.2$^o$ with respect to the $p$ strips.

%
%
\section{Importance of Longevity Studies}
\label{sec:ops}
\begin{figure}[h]
  \centerline{\hspace*{0.2in}
    \mbox{\includegraphics[width=4.00in]{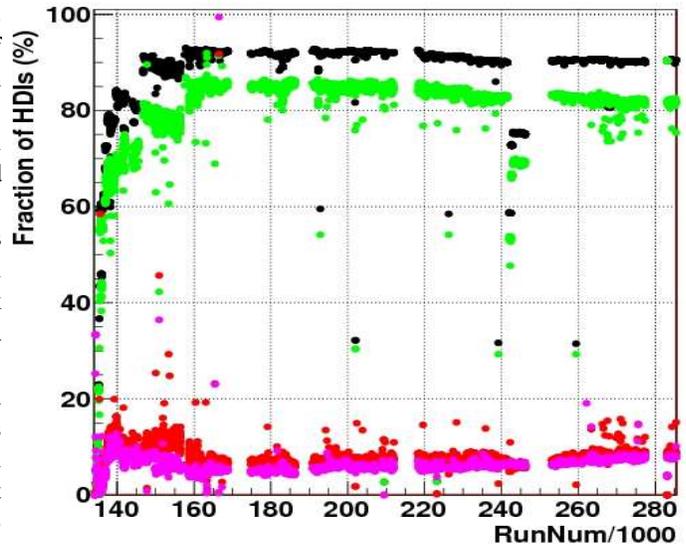}}
  }
  \caption{The fraction of integrated (black), good (green) and bad
         (red) ladders vs. run number. Shown also is the average
         digital error rate (magenta).}
  \label{fig:DigiErr}
\end{figure}

The Tevatron has delivered an integrated luminosity of 7 fb$^{-1}$
so far and is expected to run through 2010, with a possible
extension to 2011. This would expose the CDF silicon detectors to
radiation doses proportional to the collected data in excess of 9
fb$^{-1}$, far beyond their design specifications of 3 fb$^{-1}$.
In addition, aging electronics and cooling system, reducing spare
pool {\it etc.} pose significant challenges for an efficient
operation of the detectors~\cite{MarkPoster}.
Figure~\ref{fig:DigiErr} shows the current fractions of integrated
(black), good (green) and bad (red) ladders vs. run number.
A ladder is considered good if the data acquired by it
suffer from $<$1\% digital errors and bad otherwise. Shown also is
the average error rate (magenta). Over the past 8 years of
operation, the good ladder fraction has reduced by $~$3\%,
indicating an excellent overall performance since the
commissioning.

The Tevatron Run II program was originally expected to run till
2009 and accumulate 15 fb$^{-1}$ of data. This would have required
replacement of the inner layers of the CDF silicon detectors. A
revised Tevatron luminosity profile projected a much lower target
luminosity of 4-8 fb$^{-1}$ by 2009 and led to the cancellation
of the CDF Run IIb silicon upgrade project~\cite{RunIIb} in
September 2003. In 2004 a panel of CDF silicon experts made a set
of recommendations to improve the detector longevity. They include
running the detector significantly colder, implement fixes to some
identified DAQ channel failure modes, safeguard against Tevatron
beam incidents and regularly monitoring changes in sensor
depletion voltage. These recommendations were carried out
promptly. Nevertheless the radiation hardness of the sensors
remains the single most important issue for future operations.

%
%
\section{Effects of Radiation Damage}
\label{sec:rad}
The CDF silicon detector is located in  an intense radiation
field, dominated by $p\bar{p}$ collisions as opposed to beam
losses. Measurement of the ionizing radiation dose in the CDF
detector hall was carried out early in Run II~\cite{TLDs} by means
of over 1000 thermal luminescent dosimeters (TLDs) spread over
the entire detector volume. An extrapolation of the measured dose
using an inverse power law in radius gives an estimate for the
ionizing radiation dose of 300~$\pm$~60 kRad/fb$^{-1}$ at a
radius of 3 cm with a 20\% variation over $z$. This means that
L00 and Layer-0 of SVX-II have already received doses of $\sim$7
MRad and $\sim$2.8 MRad, respectively, after 7 fb$^{-1}$ of Run
II luminosity. The flux to luminosity factor has been
measured~\cite{BiasCurrent} using the increase in bias currents
of the sensors, resulting in 0.93~$\pm$~0.26 (10$^{13}$ 1 MeV n
equiv.)/cm$^2$/fb$^{-1}$ for the Layer-0 of SVX-II. The inner
sensors of L00 receive 2.5 times higher flux. The integrated
fluence in L00 after 8 fb$^{-1}$ will be of the order of what is
expected for the LHC strip trackers.

Radiation damage to the sensor bulk causes the number of
effective charge carriers to change, gradually decreasing the
voltage required to deplete the sensor until type inversion,
after which the depletion voltage increases.  The Layer-0 of
SVX-II is of most concern for this type of damage because it is
located at a radius of 2.5 cm from the beam-line and its maximum
bias voltage is limited to 170 V, as mentioned in
Section~\ref{sec:det}. Defects in silicon crystals due to
radiation damage lead to an increase of the sensor leakage
current and, as a consequence, to an increase of sensor shot
noise. In addition, signal strength degrades as crystal defects
cause degradation of charge collection efficiency. These two
effects, in combination with increasing readout ASIC noise, cause
a degradation of the signal-to-noise ratio ({\it S/N}). Therefore,
two major concerns for the longevity of the CDF silicon detector
are the ability to fully deplete the silicon sensors and
degradation of the signal-to-noise ratio.

The performance of the detector is not expected to change
significantly due to degradation of the readout electronics by
the integrated radiation dose.
Of particular concern are the readout ASIC and
the Dense Optical Interface Module (DOIM) which transmit optical
data to the front-end electronics crates. The noise in the
readout ASIC is predicted to increase by 17\%~\cite{SVX3d} and the
light output from the DOIMs to drop by 10\%~\cite{DOIM} after 8
fb$^{-1}$. Other radiation induced effects manifest themselves
as increase in Single Event Upset (SEU) rates in the front-end
electronics and power supply crates located in the CDF detector
hall. So far there has been no alarming signs of increase in
these rates.

%
%
\section{Results and Discussion}
\label{sec:results}
\begin{figure}
  \centerline{\hspace*{0.2in}
    \mbox{\includegraphics[width=3.80in]{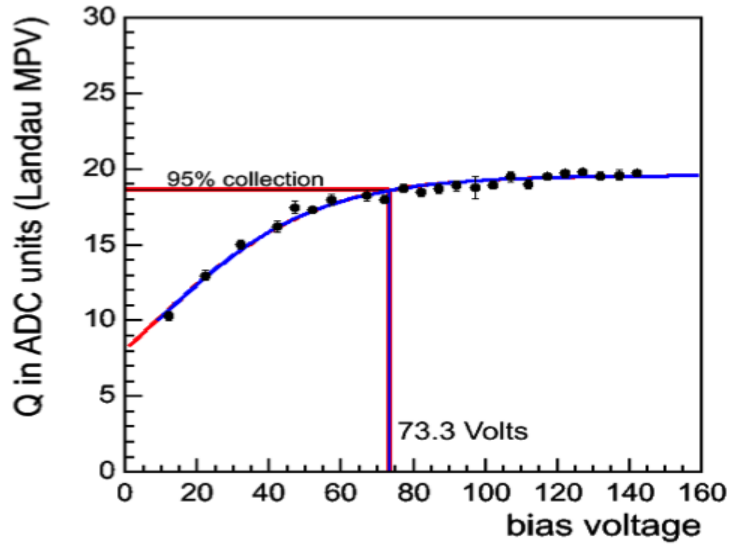}}
  }
  \caption{Determination of the depletion voltage for a ladder.
         It is defined as the bias voltage at which 95\% of
         the charge collection plateau is reached.}
  \label{fig:Vdep}
\end{figure}

\begin{figure}
  \centerline{\hspace*{0.2in}
    \mbox{\includegraphics[width=4.00in]{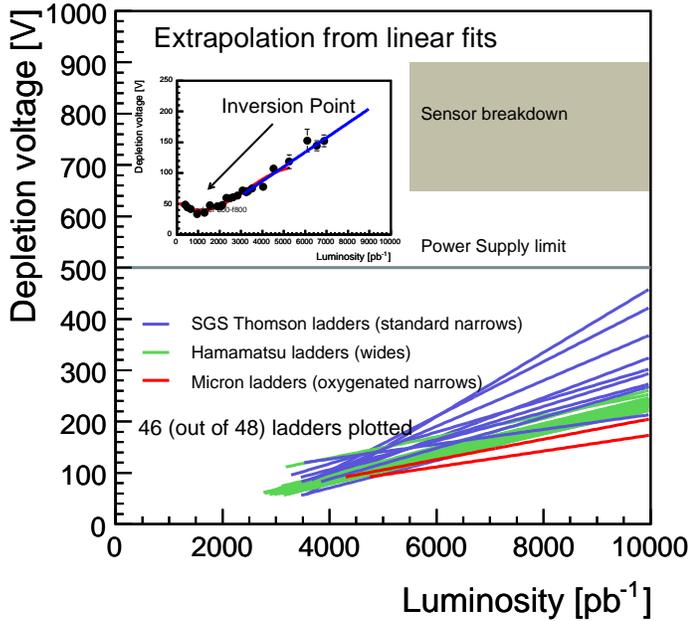}}
  }
  \caption{Prediction of L00 sensor depletion voltage up to 10
         fb$^{-1}$. The 3 makes of the sensors are shown in
         different colors. Shown also are power supply limit
         of 500 V and the sensor breakdown region.}
  \label{fig:VdL00}
\end{figure}

\begin{figure}
  \centerline{\hspace*{0.2in}
    \mbox{\includegraphics[width=4.00in]{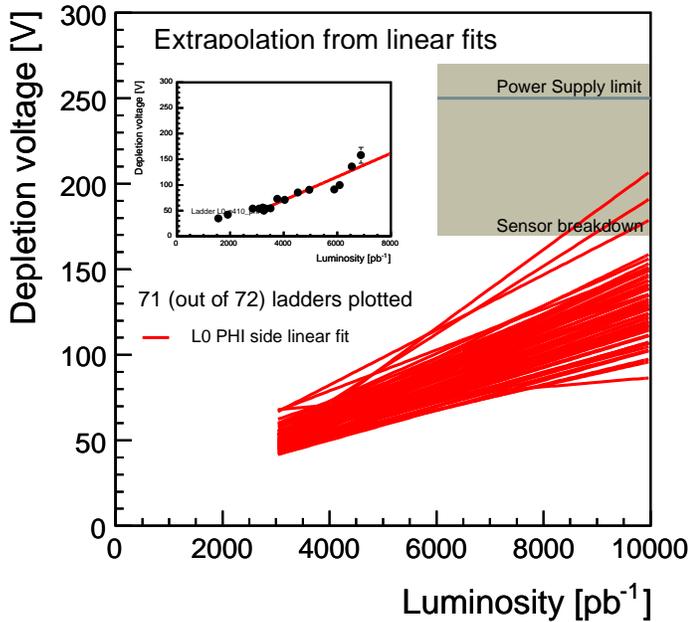}}
  }
  \caption{Prediction of SVX-II Layer-0 sensor depletion voltage
         up to 10 fb$^{-1}$. Shown also are power supply limit
         of 250 V and the sensor breakdown region (starts from
         $\sim$170 V).}
  \label{fig:VdSvxL0}
\end{figure}

For a given ladder the depletion voltage is measured in the
following way. The charge distribution is fitted to a Landau
function convoluted with a Gaussian resolution function. The
resulting Most Probable Value (MPV) is plotted against a range
of applied bias voltages, as shown in Figure~\ref{fig:Vdep}.
This distribution rises with bias
voltage and saturates to form a plateau at higher voltages. The
depletion voltage is defined as the bias voltage that collects
95\% of the charge at the plateau. The time evolution of depletion
voltage is studied as a function of the integrated luminosity. A
3$^{rd}$ order polynomial fit is performed to extract the sensor
substrate type inversion point. A linear fit is applied to data
points beyond +1 fb$^{-1}$ from the inversion point to obtain
future predictions. Figures~\ref{fig:VdL00} and \ref{fig:VdSvxL0}
show the evolution of the depletion voltage for L00 and SVX-II
Layer-0, respectively.
The positive slope of the depletion voltage evolution indicates
that all the CDF Run II silicon sensors have passed their type
inversion. Due to our condition for the linear fits, the inversion
point is roughly 1 fb$^{-1}$ below the starting point of the
lines. It is interesting to note that the CMS-type oxygenated
Micron L00 ladders (shown in red) invert much later compared to
the other two, indicating their superior radiation hardness. It
is clear from these plots that we will be able to fully deplete
most of the L00 and SVX-II Layer-0 sensors until 10 fb$^{-1}$.

\begin{figure}
  \centerline{\hspace*{0.2in}
    \mbox{\includegraphics[width=4.00in]{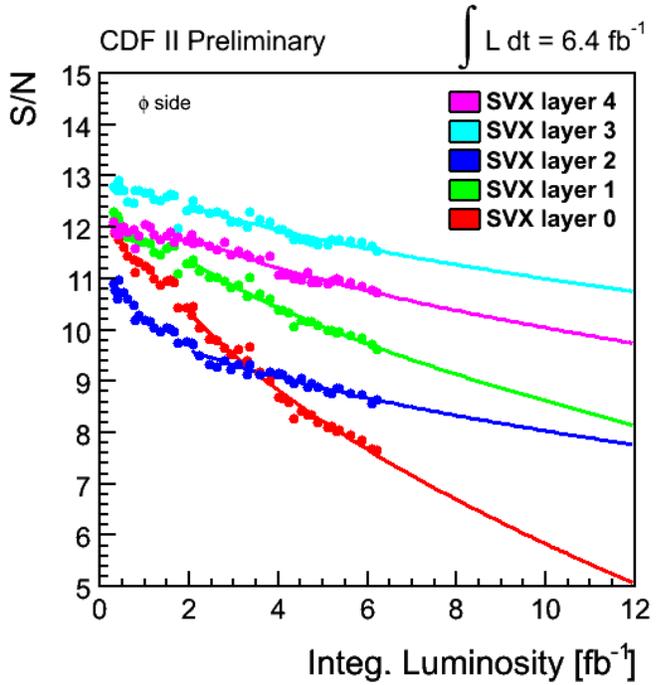}}
  }
  \caption{Prediction of {\it S/N} for the $r-\phi$ strips of SVX-II
         up to 12 fb$^{-1}$.}
  \label{fig:SNphi}
\end{figure}

\begin{figure}
  \centerline{\hspace*{0.2in}
    \mbox{\includegraphics[width=4.00in]{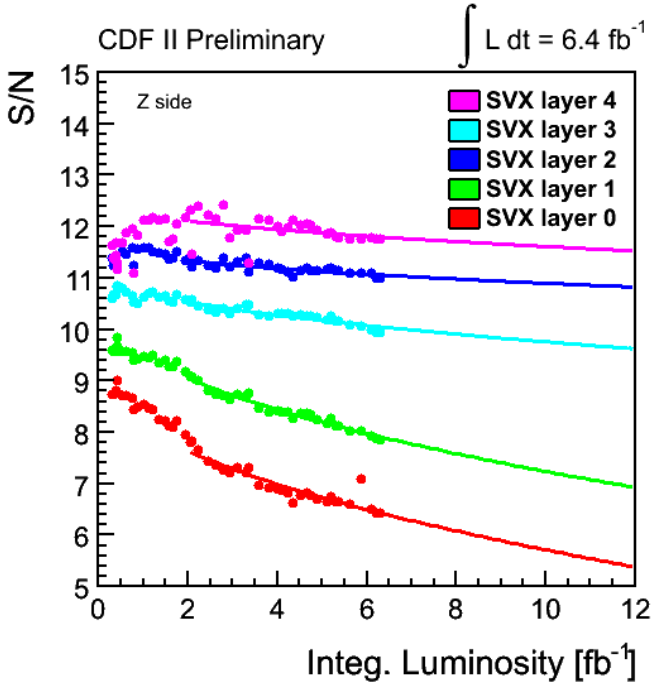}}
  }
  \caption{Prediction of {\it S/N} for the $z$ (stereo) strips of
         SVX-II up to 12 fb$^{-1}$.}
  \label{fig:SNz}
\end{figure}

For a given side, $r-\phi$ or $z$, and layer of SVX-II the
signal-to-noise ratio ({\it S/N}) is measured in the following
way. The signal, $S$, is measured as the charge collected from
hits on muon tracks reconstructed in $J/\psi \rightarrow \mu^+
\mu^-$ events collected using CDF di-muon trigger. The extracted
signal from a Landau fit is path corrected and decreases linearly
with integrated luminosity. The noise, $N$, averaged over charge
clusters in a ladder is estimated using special calibration runs
with beam. It increases as a square-root with the integrated
luminosity. Figures~\ref{fig:SNphi} and \ref{fig:SNz} show the
evolution of {\it S/N} for the $r-\phi$ and $z$ strips of SVX-II,
respectively, up to 12 fb$^{-1}$.
The plots show the predicted {\it S/N} ratios above 5 up to an
integrated luminosity of 10 fb$^{-1}$. This would correspond to
a tracking performance adequate for $b$-tagging, based on CDF
Run I experience, and for using SVX-II (especially Layer-0) in
the SVT hardware trigger~\cite{SNpred} for the remainder of Run II.

\section{Conclusions}
\label{sec:conc}
Radiation aging studies show the CDF Run II silicon detectors
are in good health after 8 years of operation. Studies of
depletion voltage and {\it S/N} ratio evolution have been
instrumental in maintaining optimal detector performance through
bias voltage upgrades. The L00 and Layer-0 of SVX-II have long
progressed through type inversion and exhibit consistent
post-inversion behavior. Most SVX-II Layer-0 ladders would be
fully depleted up to 10 fb$^{-1}$. The innermost layer, L00, is
expected to compensate for any lost tracking and vertexing
performance. In summary, the CDF Run II silicon detector will
continue successful operation for the rest of Run II, beyond 2010. \\

\enlargethispage{-1.3in}

\newpage
%
%

\end{document}